\documentclass[a4paper, 11pt]{article} 

\usepackage{amssymb, amsthm}
\usepackage{hyperref}
\usepackage{authblk}
\usepackage{verbatim}
\usepackage{enumerate}
\usepackage{url}
\usepackage{graphicx}
\usepackage{setspace} 

\usepackage[comma,authoryear]{natbib} 
\usepackage[margin=1in]{geometry}

\begin{document}
 

 \begin{flushleft}
{\Large
\textbf{Forms of social relationships in distinct cultural settings} 
}

Maroussia Favre$^{1,\ast}$, 
Didier Sornette$^{1,2, \dagger}$ 
\\
\bf{1} ETH Z\"urich, Department of Management, Technology and Economics, Scheuchzerstrasse 7, 8092 Z\"urich, Switzerland
\\
\bf{2} Swiss Finance Institute, c/o University of Geneva, 40 blvd. Du Pont d'Arve, CH-1211 Geneva 4, Switzerland
\\
$\ast$ Corresponding author: maroussiafavre@ethz.ch
\\
$\dagger$ dsornette@ethz.ch
\end{flushleft}


\section*{Abstract}
We contribute to the understanding of social relationships within cultural contexts by proposing a connection between a social theory, relational models theory (RMT: \citealp{FiskeBook, Fiske92}) and a social and political one, cultural or plural rationality theory (PRT: \citealp{Douglas78,Douglas82CB, CulturalTheory90}). Drawing examples from the literature of both theories, we argue that each relational model of RMT may be implemented in ways compatible with each cultural bias of PRT. A cultural bias restrains the range of congruent implementations of relational models, but does not preclude any relational model altogether. This stands in contrast to earlier reconciliation attempts between PRT and RMT. Based on hypothetical one-to-one mappings, these attempts expect each cultural setting to be significantly associated with some, but not all, relational models. The framework we develop helps explain the findings of these previous attempts and provides insights into empirical research by clarifying which associations to expect between relationships and cultural contexts. We discuss the theoretical basis of our framework, including the idea that RMT and PRT apply to different levels of analysis: RMT's relational models are tied to relationships between two actors and PRT's cultural biases to structures of social networks.
 
 
\vspace{0.3cm}

{\bf Keywords:} social relationships, culture, social networks, relational models theory, plural rationality theory, cultural grid-group theory

{\bf JEL classifications:} D71, A13, Z13, M14, L14, F51

\section{Introduction}

\subsection{Motivation} \label{S-motivation}

We propose a connection between relational models theory (RMT:  \citealp{FiskeBook, Fiske92}) and plural rationality theory (PRT: \citealp{Douglas78, Douglas82CB, CulturalTheory90}). These two theories connect to the works of social and political theorists and anthropologists such as Durkheim, Weber, Piaget, Marx, Polanyi, Sahlins, Evans-Pritchard, L\'evi-Strauss, and a number of others, and can be called theories of \emph{constrained relativism}\footnote{The tradition of constrained relativism goes back to Maine, Tonnies, Durkheim, and Weber. Other theories that can be said to belong to constrained relativism are, for example, Jonathan Haidt's moral foundations theory \citep{Haidt07}, Hofstede's cultural dimensions \citep{Hofstede84}, Triandis' culture orientation scales \citep{Triandis95}, Richard Shweder's tripartite theory of morality \citep{Schwederetal97} and Pierre Bourdieu's theory of practice \citep{Bourdieu77}. Except for Triandis' constructs, to which we come back in section \ref{S-studies}, the description of these theories is beyond the scope of the present work.} (\citealp{Verweij07}, \citealp[176]{LockhartFranzwa94}). 

The idea behind constrained relativism is that the variety of social life emerges from just a small number of fundamental options, and that the particular choice of an option may be largely influenced by the cultural context. This contrasts with the post-structuralist view, which holds that social life derives from an infinite diversity of options, such that each person is unique and only individual descriptions make sense. Constrained relativism also departs from rational choice theory, insofar as the latter posits that people behave as if they had access to and processed all information instantly, and used this ability to form preferences obeying well-defined axioms.

Recently, \citet{Verweij15Neuro} argued that rational choice theory, post-structuralism, and additionally behavioral economics, are inconsistent with the way our brain functions according to processes described by affective and social neuroscience. These authors propose that PRT and generally theories of constrained relativism are compatible with the findings of neuroscience. By reconciling two theories of constrained relativism, our hope is to contribute to strengthening it.

When one is confronted with the PRT and RMT typologies (to which an introduction is given in sections \ref{S-RMT-intro} and \ref{S-PRT-intro}), it immediately appears that they should be related. One may intuitively imagine one to be a simple transformation of the other. Indeed, RMT and PRT seem to describe the same fundamental ingredients of social life. PRT posits that there are four ways of organizing social relationships, each of which is associated with a compatible set of individual norms, values, perceptions, beliefs and preferences (called \emph{cultural biases}). RMT postulates four models of relationships, the \emph{relational models}, and recently examined their combinatorics \citep{Fiske2012} as well as the moral norms and values associated with them \citep{RaiFiske11, FiskeRaiVV}. Moreover, each theory claims its fourfold typology to be exhaustive and universal: in each and every social domain, people are supposed to attend to the types defined by the theory, singly or in combination.
 
As a result of the similarities between these two important theories, several researchers have previously attempted to unite them, or one of them and concepts present in the other. These efforts, which we review in section \ref{S-studies}, face related issues. \citet{Vodosek09} and \citet{Realo04} seek associations between RMT's typology and the typology introduced by \cite{Triandis95,Triandis96}, which can be argued to closely parallel the PRT typology \citep[87-9]{Verweij14Disagree}. \citet{Vodosek09} tests hypotheses expressed by \citet[50-1]{Triandis95} connecting relational models with Triandis' constructs, but Vodosek's empirical results only partly confirm these hypotheses. \citet{Realo04}, for their part, conclude that there is no one-to-one mapping between the typology of RMT and social contexts in general, but do not propose any other model. \citet{Verweij07} undertakes a reconciliation attempt of PRT and RMT at a theoretical level, which brings out several apparent inconsistencies. Finally, \citet{Brito11} empirically test for a subset of associations between relational models and social contexts, but without constructing a complete theoretical explanation for their results. 

Our approach not only reconciles RMT with PRT, but also provides interpretations for the results of the aforementioned earlier attempts, as we argue in section \ref{S-studies}. Before presenting our framework in section \ref{S-biases-RMs}, let us give an introduction to both theories. 

\subsection{Relational models theory (RMT)} \label{S-RMT-intro}

Relational models theory was introduced by \cite{FiskeBook, Fiske92} in the field of anthropology to study how people construct their social relationships. RMT posits that people use four elementary models to organize most aspects of most social interactions in all societies. These models are communal sharing (CS), authority ranking (AR), equality matching (EM), and market pricing (MP). 

\begin{itemize}
\item In CS, people perceive in-group members as equivalent and undifferentiated. CS relationships are based on principles of unity, identity, conformity and undifferentiated sharing of resources. Decision-making is achieved through consensus. CS is typically manifested in close family or friendship bonds, teams, nationalities, ethnicities or between soldiers exposed to stressful conditions.

\item In AR relationships, people are asymmetrically ranked in a linear hierarchy. Subordinates are expected to defer, respect and obey high-rankers, who take precedence. Conversely, superiors protect and lead low-rankers. Subordinates are thus not exploited and also benefit from the relationship. Resources are distributed according to ranks and decision-making follows a top-down chain of command.

\item EM relationships are based on a principle of equal balance and one-to-one reciprocity. Salient EM manifestations are turn-taking, democratic voting (one person, one vote), in-kind reciprocity, coin flipping, distribution of equal shares, and tit-for-tat retaliation.

\item MP is based on a principle of proportionality. Relationships are organized with reference to socially meaningful ratios and rates, such as prices, cost-benefit analyses or time optimization. Rewards and punishment are proportional to merit. Abstract symbols, typically money, are used to represent relative values. MP relationships are not necessarily individualistic; for instance, utilitarian judgments seeking the greatest good for the greatest number are manifestations of MP.
\end{itemize}

The four relational models (RMs) have in common that they suppose a coordination between individuals with reference to a shared model. To these, Fiske adds two limiting cases that do not involve any other-regarding abilities or coordination \citep[19-20]{FiskeBook}:
\begin{itemize}
\item In asocial interactions, individuals exploit others and treat them as animate objects or means to an end (as in psychopathy, armed robbery, pillage);
\item In null interactions, people do not interact at all (they do not actively ignore each other, which still requires a coordination), as can be the case of two inhabitants of the same building who never cross each other's way or fail to notice each other's existence, and thus do not adapt their actions to each other.
\end{itemize}

Any relationship generally consist in a combination of RMs \citep[155-168]{FiskeBook}. For example, ``roommates may share tapes and records freely with each other (CS), work on a task at which one is an expert and imperiously directs the other (AR), divide equally the cost of gas on a trip (EM), and transfer a bicycle from one to the other for a price determined by its utility or exchange value (MP)" \citep[711]{Fiske92}. Or, ``any aspect of the relations between husband and wife can be structured as CS, AR, EM, MP, or any combination of these models across different domains of the marital relationship" (ibid., 712).

The RMs can be seen as ``templates" \citep[711]{Fiske92} or ``structures" (ibid., 690) with ``flexible application" (ibid., 692). RMs, being abstract structures, can be applied in many different ways to any aspect of any social domain. Cultures have their own values or expectations regarding when and how to execute each RM. These cultural paradigms are called ``preos" by \citet{Fiske2000}. For example, EM can be implemented under the form of equality of treatment, results or opportunity. In different applications of MP, rewards may be proportional to productivity, effort or ability \citep[712-3]{Fiske92}. AR relationships may be implemented based on age, education or gender, and can take the form of calling someone by last instead of first name, following orders, or giving precedence. CS expressions include sharing comestibles, dancing in unison, or body contact and sex \citep[64]{FiskeConstitMed04}. These examples are far from exhaustive: cultural diversity may emerge from each culture specifying differently how, when and where each RM applies \citep[77 and 79]{Fiske2000}.

The majority of examples of RMs given by \cite{FiskeBook} from various societies around the world are of interactions between two people, and sometimes between two groups. Yet, Fiske also uses RMs to characterize single groups of more than two individuals in which all members use the same relational model in the context of a social activity. For example, if members of a group are all implementing CS when sharing food, it can be called a ``CS group" with respect to that activity \citep[151]{FiskeBook}. Rotating credit associations \citep[153]{FiskeBook} or equal distribution of any common resource are typical examples of EM within a group of more than two people. In such situations of homogeneous collective action, a single RM gives an accurate description of what happens between any two members and thus can be used to characterize the group as a whole. That is, a CS group can be entirely described by an ensemble of dyadic CS relationships.

RMT has motivated a considerable amount of research that supports, develops or applies the theory, not only in its original field of social cognition \citep{Fiskeetal91, Fiske93, Fiske95, FiskeHaslam97}, but also in diverse disciplines such as neuroscience \citep{Iacoboni04}, psychopathology \citep{Haslametal02}, ethnography \citep{Whitehead02}, experimental psychology \citep{Brodbecketal13}, evolutionary social psychology \citep{Haslam97}, and perceptions of justice \citep{Goodnow98}, to name a few. \cite{Haslam04}, \cite{RMT_overview} and \cite{Whitehead93} provide reviews of this research. Moreover, as suggested by \citet[210-23]{FiskeBook}, the RMs map onto the four measuring scales defined by \citet{Stevens46}, and their symmetry properties evoke analogous features of neural activity \citep{Bolender08, Bolender}.

\citet{FavreSornette15} recently introduced a model of social relationships rooted in the observation that each agent (individual or group) in a dyadic interaction can do either the same thing as the other agent, a different thing, or nothing at all. We showed that the relationships generated by this model aggregate into six exhaustive and mutually disjoint categories of interactions. We argued that these six categories offer suitable abstract representations of the social actions performed by agents implementing the four relational models, and the asocial and null interactions. This representation of social relationships suggests that the RMs form an exhaustive set of all coordinated social relationships. 

\subsection{Plural rationality theory (PRT)} \label{S-PRT-intro}

Plural rationality theory (PRT) was initiated by \cite{Douglas78,Douglas82CB} and developed further by several other researchers, notably \cite{CulturalTheory90}. For a review, see \citet{Mamadouh99}. Different authors have given this theory different names, of which cultural theory, grid-group analysis (e.g. \citealp[220]{KingCoyle94}), the theory of socio-cultural viability \citep[15:n5]{CulturalTheory90}, neo-Durkheimian theory \citep{Perri03}, and plural rationality theory \citep{Verweij15Neuro}. The hypotheses at the basis of PRT are summarized by \cite{Verweij11HowToTest} and \citet[35-8]{ClumsyVerweij11}; here we follow this summary but select only the points that we need for our present purposes.

PRT posits that there are only four viable ways of organizing social relations: hierarchy, egalitarianism, individualism and fatalism. Two dimensions underlie this typology: group and grid.

\begin{itemize}
\item Group (collectivity, integration) quantifies the extent to which people are incorporated into a larger social unit, fostering group solidarity and group pressure;
\item Grid (stratification, regulation) measures the degree of prescribed asymmetry between people, promoting obligations constructed by roles. 
\end{itemize}

Each dimension is assigned high and low values, which results in four patterns of social relations: hierarchy is high group, high grid; egalitarianism is high group, low grid; individualism is low group, low grid; and fatalism is low group, high grid. 

PRT further claims that from each pattern of social relationships emerges a compatible so-called cultural bias that specifies, very generally, ways of perceiving, justifying, reasoning, and feeling. This includes perceptions of nature, human nature, time, risk, leadership, justice, blame, and governance. Each cultural bias, in turn, includes values that support and justify its corresponding pattern of social relationships. The idea that each bias comes with a rationality of its own is at the root of the name ``plural rational theory." The salient features of the cultural biases are as follows.

\begin{itemize}
\item In a hierarchy (high grid, high group), actors relate to each other based on their prescribed relative positions. Superiors and inferiors share ethical values, such as respect for other members, and identify themselves with the collective. Nature, and generally the world, are seen as controllable, and stable within limits that can be determined by certified experts. Perception of time is long-term, supporting thorough planning. Human nature is seen as sinful and thus calling for adequate regulation.
\item In egalitarianism (low grid, high group), actors insist that people must start and end up equal. In fact, this principle extends to all living things, including animals and plants. Nature is seen as fragile and intricately interconnected. The earth must be taken care of in the sake of future generations. Solutions to social problems must be urgently implemented to avoid global disasters. Human nature is thought of as fundamentally altruistic, but subject to corruption by status and power. 
\item In individualism (low grid, low group), actors are and see others as mostly self-interested, without this being seen as a fundamentally bad thing. While egalitarians pursue equality, individualists pursue liberty \citep[223]{KingCoyle94}. Efficiency, autonomy, individual achievement, skill-based task distribution, entrepreneurial spirit are valued. Accordingly, competitive markets should not be constrained by bureaucratic regulations. Seen as another autonomous actor, nature itself is expected to be resilient, recovering equilibrium on its own upon external perturbations. 
\item In fatalism (high grid, low group), actors are confined in prescribed asymmetrical roles, as in a hierarchy. But since solidarity is virtually inexistent, actors mostly fend for themselves. Manipulative, unpredictable, deceitful despots are free to exploit a society of isolated individuals whose mutual distrust and amoralism prevents them from standing together. Nature is seen as another unknowable and random actor; time never brings any fundamental change. According to fatalism, there is no meaningful, reliable pattern to be found in anything.
\end{itemize}

According to PRT's ``multiple selves" hypothesis \citep[265-67]{CulturalTheory90}, people may not fit neatly into the same box in all respects. Instead, they may prefer, or live under, the influence of different cultural biases in different domains of their social life. The PRT literature emphasizes that cultural biases are not personality types (e.g. \citealp[19]{Thompson08-OD}, \citealp[106-8]{RaynerIn92}). In the present work, we talk about hierarchical, egalitarian, individualistic and fatalistic actors. These terms do not refer to distinct types of people, but instead to individuals or institutions following the principles of a given cultural bias in a given respect. We talk about settings (e.g. an egalitarian setting) to denote social contexts organized along the lines of a grid-group pattern and composed of individuals presumably behaving according to the corresponding cultural bias.

Finally, while the four cultural biases stand in contradistinction to each other, PRT predicts that attempts to resolve social problems (from urban planning to international conflicts to global warming, going through the whole spectrum of human activities) that do not address the views and goals of all four cultural biases are bound to fail, not only according to the proponents of the neglected biases, but also on their own terms. PRT thus gives a sure recipe for failure, but not a sure recipe for success - taking into consideration all ways of life is a necessary, but not sufficient, condition for success. An important and growing part of the research in PRT consists in case studies documenting failed or successful attempts to resolve social problems and verifying the latter prediction (e.g. \citealp{CoyleEllis94,ClumsyEds06, ClumsyVerweij11}). In the present work, we support our thesis by drawing examples from this literature as well as from the RMT literature.

\subsection{Aim and scope}

It may be helpful to consider our approach in the perspective of earlier attempts to connect RMT with PRT. These typically assume that high group is associated with CS; low group with MP; high grid with AR; and low grid with EM (\citealp[50-1]{Triandis95}, \citealp{Triandis98,Vodosek09}). Accordingly, the combination of high group and high grid (the hierarchical bias) is expected to be associated with CS and AR, but not with other RMs; egalitarianism, with CS and EM; individualism, with EM and MP, and fatalism with AR and MP. As mentioned in section \ref{S-motivation} and reviewed in section \ref{S-studies}, this and similar one-to-one mappings have brought only partly satisfactory results \citep{Realo04, Verweij07, Vodosek09, Brito11}. 

In contrast to these attempts, we propose that there is no one-to-one correspondence between the RMs and PRT's grid and group dimensions, or the cultural biases. Instead, we argue that each cultural bias is compatible with any RM, but not with all of its implementations. Differently said, all RMs can be used to constitute social relationships compatible with a cultural bias. Importantly for empirical research, this means that an association may be found between each RM and any cultural bias. 

This view builds on the idea that the implementation of an RM is specified by cultural rules. As \citet[713]{Fiske92} puts it, ``each of the four universal models can be realized only in some culture-specific manner; there are no culture-free implementations of the models. [...] The models cannot be operationalized without specifying application rules determining when and to whom and with regard to what they apply [...]. A major problem for future researchers is the study of how people select a particular implementation of a specific model in any given context. People are probably guided primarily by cultural rules [...]." Our thesis consists in arguing that these cultural implementation rules may be specified, to a certain extent, by the cultural biases of PRT.

In this chapter, our main focus is on how a cultural bias specifies which implementations (or expressions) of each RM may be accepted, preferred, customary or obligatory in social settings associated with this cultural bias. However, the reverse mechanism may also be considered. Namely, particular RMs implementations may evoke, support or constitute a social setting associated with a cultural bias compatible with these implementations. We discuss how to approach this reverse mechanism in section \ref{S-bottom-up}, but examining it in details is beyond the scope of this work.

In section \ref{S-biases-RMs}, we examine how each relational model may be implemented in ways compatible with each cultural bias. We illustrate our points with cases drawn from the RMT and PRT literature. 

In section \ref{discussion}, we clarify the theoretical hypothesis underlying our framework. Namely, our view is that relational models are tied to relationships between two actors (persons, or groups), and cultural biases to social networks with characteristic structures. We examine this hypothesis in the context of RMT's recent extensions introducing metarelational models \citep{Fiske2012} and moral motives \citep{RaiFiske11, FiskeRaiVV}. We connect this discussion to the question of how our framework could be built in the other direction, that is, by specifying how specific RM implementations constitute congruent cultural biases.

In section \ref{S-studies}, we review previous studies \citep{Realo04, Verweij07, Vodosek09, Brito11} manifesting ambitions similar to ours and interpret their results in the light of our framework. These considerations lead us to suggest directions for empirical research. 

\section{Relational models within each cultural bias} \label{S-biases-RMs}

In this section, we consider the four cultural biases in turn and discuss how relational models may be implemented in ways compatible with each bias. We also describe how asocial  and null interactions may be perceived by actors in each bias. We consider first the interaction modes that are most salient, so that the order is different within each bias. 

\subsection{Hierarchy (high group, high grid)} \label{S-biases-H}

The correspondence between the PRT hierarchical ideology and AR relationships is very apparent. A common feature of hierarchical settings and AR relationships is that superiors abide by moral principles and subordinates are not oppressed. According to \citet[14]{FiskeBook}, in AR, ``Followers are entitled to receive protection, aid, and support from their leaders;" AR ``contrasts with pure coercive power [...] subordinates believe that their subordination is legitimate." According to a principle of \emph{noblesse oblige}, ``authorities have an obligation to be generous [...] and exhibit pastoral responsibility in protecting and sustaining their subordinates" \citep[700]{Fiske92}. In PRT's hierarchical bias, authority is also legitimate, and the hierarchical institution is intended to benefit all of its members. ``The top level of authority must never fail to respect the lowest" \citep[98]{Douglas05};  ``hierarchists believe that human beings are born sinful but can be redeemed by good institutions" \citep[35]{CulturalTheory90}; the domestic governance ideal of the hierarchical bias is ``wise, benevolent" \citep[56]{ClumsyVerweij11}.

Let us now argue that MP (i.e., a relational structure oriented toward the principle of proportionality) is an interaction mode that also suits well the hierarchical bias. This is because MP is particularly convenient for bureaucratic top-down allocation, based on the recommendations of experts recognized as holding authority, and following well-defined rules. In fact, a canonical example of MP in RMT is ``bureaucratic cost-effectiveness standards (resource allocation as MP)" (\url{http://www.rmt.ucla.edu}, last accessed 2016-03-04). A manifestation of MP is the constitution of a group defined as a ``bureaucracy with regulations oriented to pragmatic efficiency" \citep[695]{Fiske92}.

An MP-type of thinking is associated with hierarchical actors in numerous case studies carried out in PRT. For example, in the context of the Kyoto protocol intended to address global warming, tradable carbon credits is an MP-type instrument introduced by an institution identified as highly hierarchical, the United Nations Framework Convention on Climate Change (UNFCCC) \citep[40-69]{ClumsyVerweij11}. Instituting quotas on ethnic minorities in Dutch schools, sport clubs and housing is interpreted as a ``quintessentially hierarchical undertaking" by \citet[99]{DutchInClumsy06}. In the context of seat belts legislation, ``the hierarchical actors believed that a law would save large numbers of lives [...], conveniently rounded to 1000 lives [...] a year [...] and produced abundant [statistical] evidence to support the belief" \citep[148; see also 152]{SeatBeltsInClumsy06}. Finally, the problem of car mobility ``has two sides [...]: a demand side (the need for mobility) and a supply side (the infrastructure capacity). [...] These are problems that can be handled in a technocratic way [by a hierarchical actor]" \citep[54]{CarsInCoyle}.

Hierarchy is high group, which ``requires a long-term commitment and a tight identification of members with one another as a corporate identity," and makes ``the unit more exclusive and conscious of its boundary" \citep[36 and 35]{ClumsyVerweij11}. This sense of shared belonging to a larger, bounded whole is also at the basis of CS: ``people attend to group membership and have a sense of common identity [...] what is salient is the superordinate group as such, membership in it, and the boundaries with contrasting outsiders;" people ``identify with the collectivity" \citep[13-4]{FiskeBook}. In that sense, CS relationships support and are supported by the notion of high group in general, and may operationalize this fundamental feature of hierarchy in particular. 

Yet, high group and high grid also mean that group pressure and prescribed roles constrain social relationships. In a hierarchical setting, CS relationships thus have to be respectful of and determined by the hierarchy. Furthermore, collective rules (which can change from one hierarchical group to another) determine who can (or must) engage in CS with whom, and which specific implementations thereof are allowed. CS relationships may thus be constituted through rituals that emphasize and are determined by a common identity with the hierarchical system. \citet[168-193]{Dumont-HH} gives detailed descriptions of such contact and food rituals in the Indian caste system. Regulating who may or may not have sexual relations is another example of hierarchical intervention in the domain of CS relationships.

Similarly to CS, EM is implemented only with the consent or even under the imposition of the hierarchy. \citet[96, 98, 100]{Douglas05} insists that competition is restricted in a hierarchy. In this context, EM relationships can be particularly adequate to foster cooperation in specific domains of social life: giving a few expressions of EM, \citet[704]{Fiske92} notes that each of them ``simultaneously builds group (or dyadic) solidarity." For example, \citet[99]{Douglas05} recounts that in her grandparents' hierarchical home, ``No competition was allowed between my sister and myself [...] the general rule was equality between us; we were expected to share presents." 



In a hierarchical setting, asocial interactions may be seen as a violation of formal rules of conduct, and may be punished so as to reinforce rules adhesion. This is illustrated by the case of the Internet, where the (very few) hierarchical actors present in this setting ``focus more explicitly on things like content-regulation (censorship laws), restrictions on the use of encryption tools, enforcement of copyright laws, ways of combating crackers, apprehending terrorists, defeating drug dealers and so on" \citep[208]{WebInClumsy06}. Moreover, according to actors identified as hierarchical by \citet[113]{ClumsyVerweij11} in the context of the invasion of Iraq by the United States in 2003, ``Threats to the international order [...] should be dealt with [...] by scrupulously following international law." 

Null interactions are not viable within a hierarchy. As \citet[226]{Douglas92RiskBlame} writes, ``Since no one can be eliminated, all have to be assigned places in the system [...] people can drop down, but not out.'' Hence, the exclusion of a member (which is an imposed null interaction) is likely to be a last resort. In general, null interactions can take the form of indifference toward out-members and operationalize the boundaries of a hierarchy. 

\subsection{Egalitarianism (high group, low grid)} \label{S-biases-E}

Egalitarian settings clearly support and are supported by EM relationships. Equality between members of an egalitarian group is maintained through EM relationships. EM relationships can nearly be said to constitute the norm of egalitarian settings. There is however an expression of EM that may not be the first choice in egalitarianism, namely the one-person, one-vote standard, since egalitarian actors rather favor consensus decision making (e.g. \citealp[83]{ClumsyVerweij11}). Still, egalitarians may have to resort to the one-person, one-vote principle when reaching a consensus is not realistic (as, for example, in a large group).

Some combinations of EM may be compulsory in egalitarianism, for example equality of opportunity and outcome. Only the former will not do, as we see in section \ref{S-biases-I} that equal opportunity alone corresponds to individualism. Egalitarianism seeks to prevent any form of discrimination from the beginning to the end of any process.

CS relationships are also largely compatible with the egalitarian ideology. For example, as mentioned above, egalitarian decision-making process is by consensus \citep[83]{ClumsyVerweij11}, which is precisely the way decisions are reached in CS relationships \citep[695]{Fiske92}. Moreover, mutual aid follows the same principle in PRT's egalitarianism and CS. Indeed, communism is interpreted by PRT as corresponding to the egalitarian way of life \citep[155-7]{CulturalTheory90}. The communist maxim ``From each according to his ability, to each according to his needs" parallels CS mutual aid, also based on respective abilities and needs \citep[693]{Fiske92}. Finally, in egalitarianism, actors see ``man as essentially caring" and trust each other \citep[4]{ClumsyInClumsy06}, while CS is seen as ``generating kindness, generosity" \citep[14]{FiskeBook}; ``people have a sense of solidarity" \citep[13]{FiskeBook}. This last point is linked to high group being particularly conducive to CS relationships, as we argued in section \ref{S-biases-H}. In general, the egalitarian bias guides implementations of CS in the sense that there is a group pressure to implement specific CS relationships (high group), and these relationships are seen as appropriate or obligatory between all members indiscriminately (low grid).

Egalitarianism is mostly non-conducive to AR relationships, as the asymmetry that is both a precondition and a consequence of AR relationships fundamentally contradicts the egalitarian ideology. Yet, for a large group to reach any collective goal, some degree of stratified organization operationalized by AR relationships is often unavoidable. This is pointed out by \citet[367]{Mars08}, talking about an egalitarian group: ``any form of authority is resented", yet ``All organisations, however egalitarian, `co-operative' and enclavic they may be, none the less need a degree of order and control." In Mars' study, egalitarian actors come up with an idea that reconciles the implementation of AR relationships with egalitarianism: ``to have `rolling co-ordinators' who would each hold office for a month at a time" (369). Rotating chairmanships is mentioned by \citet[703]{Fiske92} as an EM scheme for group choice. This constitutes an interesting combination of AR and EM under the same umbrella of egalitarianism. In general, egalitarian leadership (implementing AR) is charismatic: a leader is chosen among and by her peers based on her charisma, and is followed voluntarily \citep[57 and 92]{ClumsyVerweij11}. 

Since MP offers an equity principle justifying unequal allocation (e.g. reward proportional to contribution), it can seem fully incompatible with egalitarianism. Moreover, individualistic and hierarchical actors are inclined to adopt MP-based reasonings to implement their own views (see sections \ref{S-biases-H} and \ref{S-biases-I}), which egalitarian actors outright oppose. 

However, there are expressions of MP that are actually fully compatible with egalitarian principles. These are for example so-called fair trade, and barter. Fair trade addresses inequalities of wealth between nations by attempting to help producers in developing countries. Barter (moneyless exchange) provides the opportunity to acquire second-hand, home-made or homegrown products. This is in line with egalitarian environmental concerns and preference for low consumption: ``we must all tread lightly on the Earth" \citep[37]{ClumsyVerweij11}; the egalitarian economic ideal is ``minimal and local production and consumption" (ibid., 90). An egalitarian barter system would be such that anyone can enter it (equality of opportunity) and such that everyone benefits (equality of outcome). One can find various barter systems online; these often promote mutual enrichment and community building. 

Other examples of MP expressions compatible with egalitarian principles are microcredit aimed at empowering women in developing countries, or policies according to which the rich should pay more taxes than the poor, proportionally to a reference point such as fortune or income. Cost-benefit calculations (corresponding to MP) may also be carried out within contexts favored by egalitarian actors, such as the development of renewable energies.

The main conditions on egalitarian MP thus include equal opportunities to enter deals, and equally beneficial outcomes for all parties. \citet[370]{Mars08} illustrates this idea concerning the use of an MP-type instrument in an egalitarian setting: ``To avoid drawing too obvious a parallel with [...] individualist practices, however, this re-introduction of the levy had to be cloaked in the rhetoric of egalitarianism;" in the end, the preferred solution would be such that ``Everyone would [...]  benefit."

Asocial interactions are seen as possibly pathological by egalitarian actors; criminals should be helped and oriented toward harmonious reintegration. The lengths through which actors abiding by egalitarian principles are ready to go to include asocials are documented by \citet[76-7]{SwedlowInCoyle} in the context of a mental health facility, the Kingsley Hall.

As in a hierarchical setting, the ideal of group cohesiveness renders null interactions not viable within an egalitarian group. Null interactions are imposed only under the form of last resort exclusion. Residents of the aforementioned egalitarian Kingsley Hall \citep[77]{SwedlowInCoyle}, despite feeling ``severely threatened" by the behavior of one of them, would still wonder ``what would happen to him if we told him to leave?"

\subsection{Individualism (low group, low grid)} \label{S-biases-I}

Individualism may appear to rely more insistently on MP than on any other RM. This may be largely due to MP being particularly well suited to immediate individual optimization. Indeed, MP-based bargain, contract or agreement directly between the involved parties allows each party to minimize her cost/benefit ratio. In PRT case studies, examples abound of individualist actors being equated with companies obviously functioning mainly on the basis of MP \citep{CoyleEllis94, ClumsyEds06, ClumsyVerweij11}. Indeed, MP ``is a relationship mediated by values determined by a market system" \citep[15]{FiskeBook} and companies are precisely built to function as individualist actors in a competitive market. 

However, MP relationships are not fundamentally self-interested. ``MP relationships need not be selfish, competitive, maximizing, or materialistic - any of the four models may exhibit any of these features. MP relationships are not necessarily individualistic" (\url{http://www.rmt.ucla.edu}, last accessed 2016-03-04). We argue in sections \ref{S-biases-H} and \ref{S-biases-E} that specific variations of MP are compatible with the hierarchical and egalitarian biases, which are at odds with the individualist bias. Conversely, any other interaction mode may serve self-interest, and can thus be made compatible with the individualistic bias. 

In PRT, individualism favors the principle of equal opportunity \citep[57]{ClumsyVerweij11}, an expression of EM. However, let us stress that the principle of free competition then takes over, resulting in inequalities (of wealth, for example), something that individualistic actors would accept (but not egalitarian ones). Another EM scheme, the ``one person, one vote" electoral process allows individualistic actors to vote in line with their perceived best interests. Moreover, individualistic actors may be prone to use EM in an alliance with selected peers in order to attain their personal goals. This is illustrated by the collusion of two groups in the context of the Russian transition to capitalism \citep{RussiaInClumsy06}, whose analysis suggests that it followed an extreme individualistic logic. Finally, another example of individualistic EM, inspired by \citet[224]{KingCoyle94}, is that of rock climbers taking turns belaying each other, not necessarily in order to manifest any kind of perceived equality, but simply in order to practice the sport for themselves. 

AR, like any other RM, comes in a number of variations. These include prestige rankings, as in sports team standings, and expertise-based leadership: ``roommates may [...] work on a task at which one is an expert and imperiously directs the other (AR)" \citep[711]{Fiske92}. These instantiations of AR are compatible with the individualistic ideology; prestige rankings valorize individual achievement, while following an expert's advice can serve self-interest. 

Other instantiations of AR, such as that based on the idea of a superior commanding an inferior, go against autonomy, the core value of individualism. Yet, if an individualistic actor estimates that entering such an AR relationship will be beneficial to herself, she may just do so. In fact, in individualism, concrete outcomes tend to matter more than deep principles underlying relationships. It may thus easily occur for an individualistic actor to hold a different view than the other party as to what principles the relationship fundamentally follows. As long as it is worth her while to do so, an individualistic actor may just play along the rules preferred by her counterparty, thus avoiding a time- and energy-consuming conflict. 

PRT abounds in examples of individualistic actors (typically companies) associating with hierarchical groups (typically governments). This individualism-hierarchy alliance can be so common and strong that it has been called ``the establishment" \citep[88]{CulturalTheory90}. It is illustrated by \citet[56]{CarsInCoyle}, who also uses the term ``establishment," in the context of traffic policy projects. In general, this relationship can be interpreted as AR (from the hierarchical actor's viewpoint), MP (from the individualistic actor's viewpoint), or even EM (e.g. if the actors maintain equality in some respect). An example of relationship that may be interpreted as either expertise-based AR, or MP, is that of an individualist governmental party seeking the expert advice of McKinsey on the clean-up of the Rhine \citep[193]{ClumsyVerweij11}.

Competitive contexts, such as individual sports (e.g. racing), or competition for professional status, provide instructive examples of individualistic settings. Such competitions are often seen as fair only if participants start as equals (on the same line for a race, with the same education level for professionals); this is EM. However, the process culminates in asymmetrical rankings glorifying those at the top for their achievements (AR). It may also result in the attribution of some form of reward proportional to the degree of these achievements, for example a yearly bonus (MP). These manifestations of EM, AR and MP fall under the same umbrella of a process thought of in individualistic terms. This example illustrates that thinking simultaneously in terms of relational models and cultural biases may help understand the combination of relational models in dynamical processes, and the presence of apparently contradictory values in the same cultural bias.

In an individualistic perspective, CS relationships cannot be supervised or imposed by collective principles, as in hierarchy and egalitarianism. Instead, individualists may freely constitute CS relationships based on their personal inclinations, and expect others to do the same. For example, as regards ``sex between consenting adults, individualists will be against allowing the government to intervene" \citep[66]{Dake91}. In the same spirit, a CS relationship may be given up if it stops bringing personal enjoyment or benefits. Yet, \citet[220]{Olli12}, interviewing an individualistic household, notes ``how strong the family loyalties are," which however ``is not a counter indicator of the individualistic way of organizing." Olli adds that PRT ``actually does not say that the loyalties are weak. It says that the group pressure is weak, and that the obligations constructed by roles are weak" (n171). Individualists may thus engage in loyal CS relationships, but only with selected others, and feel entitled to orient these relationships toward the satisfaction of their own wishes or needs, while giving others the same right. Conformity to and identification with a group may occur, but again only because individualists choose this for themselves.

While the above points evoke relationships between two people, CS relationships can also be constituted between individualistic groups: for example, while the acquisition of a company by another is MP, the merging itself is an expression of CS, whereby the two companies pool their resources and literally become one. At the same time, the two companies' departments may be newly positioned relatively to each another in a global organigram depicting AR relationships. The agreement between the companies may have consisted in negotiating this reorganization based on the interests and expertise of the respective departments, in an individualistic way. This is another example of a combination of RMs (MP, CS, AR) along a process globally guided by the individualistic bias.

In an individualistic setting, free riding can be seen as fair game, so that asocial interactions may be expected. Consequently, actors may find it normal to take individual measures against free riding, e.g. by owning a gun \citep{GunsInClumsy06}, contracting an insurance or hiring a lawyer. Victims of asocial interactions may be inclined to seek justice for themselves through revenge. At the same time, individualistic actors may defend asocials' freedom: indeed, they grant others the same right to freedom as themselves. \citet[81-5]{SwedlowInCoyle} reviews the individualistic view on the mentally ill, according to which people should not be committed in a mental health institution against their will, apparently no matter how dangerously asocial their sickness may make them.

Null interactions are common and do not carry the heavily loaded meaning of last resort exclusion present in hierarchy and egalitarianism. Instead they denote ordinary, voluntary withdrawal from relationships seen as unproductive. As \citet[223]{KingCoyle94} puts it, ``freedom entails the choice not to participate, be it by not voting in politics, [...] or by withdrawing from social contact altogether."

\subsection{Fatalism (low group, high grid)} \label{S-biases-F} 
 
Fatalism, being by definition unfavorable to social coordination, should actually mostly hinder the constitution of relational models. Instead, the asocial interaction may dominate social life in fatalistic settings. In what has become a classic illustration of fatalism \citep[223]{CulturalTheory90}, \citet[10]{Banfield58} introduces the concept of ``amoral familism" to describe a peasants society in southern Italy. Amoralism is at the basis of asocial interactions: indeed, moral rules are shared social constructs, and asocial interactions are characterized by the absence of any shared directive model (\citealp[20]{FiskeBook}, \citealp{RaiFiske11}). The pervasive presence of asocial interactions in fatalism is illustrated throughout Banfield's account. Another illustration of how fatalistic actors expect interactions to be asocial by default is given by \citet[135-7]{ClumsyVerweij11}. He suggests that some actors observing or involved in the Rwandan genocide interpreted it through the lens of the fatalistic cultural bias. For example, according to the (implausible) fatalistic story told by these actors, ``everybody committed genocide" (ibid., 135).

Null interactions are also very common in fatalistic settings. Fatalistic actors have also been called ``isolates" (e.g. \citealp[105]{Douglas92RiskBlame}, \citealp[114]{Douglas05}). Mitterrand, described by \citet[168:n185]{ClumsyVerweij11} as a fatalistic leader, actively promoted null interactions among his staff; ``they were not encouraged to get together (in fact they were discouraged from developing joint viewpoints)." Such imposed null interactions can ensure the submission of oppressed actors by preventing them from coordinating any kind of revolt. France's inaction in Rwanda \citep[159]{ClumsyVerweij11} can also be seen as illustrating the prevalence of null interactions in fatalism. The isolation of fatalistic actors is illustrated by \citet[246]{Olli12}: in a fatalistic household, one spouse has friends but does not invite them home and ``emphasizes the separateness of their lives," and the other has no friends. In general, null interactions sustain the status quo that constitutes the fatalistic norm; the fatalistic perception of time is ``undifferentiated: same old, same old" \citep[57]{ClumsyVerweij11}. 

Social stratification (high grid) evokes AR but, as suggested by our treatment of low-grid biases (sections \ref{S-biases-E} and \ref{S-biases-I}), we do not associate AR exclusively with high grid. In fact, a fatalistic setting is rather unfavorable to the constitution of AR relationships, because these suppose the common acceptance of social norms, a feature mostly unsupported by low-group environments. We thus suggest that in most cases, asymmetrical relationships (e.g. fatalistic leadership) actually instantiate the asocial interaction, possibly under the pretense of functioning through regular AR relationships. The possible confusion between AR and asocial is noted by \citet[702]{Fiske92}, though the other way around: ``Because Westerners often fail to distinguish between coercive force and AR as a relationship, they tend to see many AR interactions as illegitimate." 

The role of asocial interactions as substitutes for genuine AR relationships is illustrated, for example, by slave culture, which \cite{SlaveryInCoyle} argues to be fatalistic. Slavery is the result of asocial oppression, while the relationship master-slave could be falsely equated with AR, especially by the masters themselves. Also, fatalistic leadership has been qualified as ``unpredictable, secretive, ruthless, cunning" and illustrated by the governance style of French President Mitterrand \citep[57, 174]{ClumsyVerweij11}, which corresponds to the asocial interaction, and not AR. 

That being said, AR relationships may still be constituted. A superior may support her prot\'eg\'e for as long as this appears to be in her self-interest, while the willing follower benefits from a relationship she perceives as legitimate (at least temporarily). However, like other RMs constituted in a low-group setting, the relationship is not constrained or determined by collectively shared rules. Instead, it is viable only as long as both parties individually estimate it to be useful to themselves, which in fatalism can randomly cease to be the case at any time. The randomness and unpredictability of relationships in fatalism is exemplified by the fatalistic domestic governance ideal described as a ``personal rule (fiefdom); at best benevolent dictator" \citep[56]{ClumsyVerweij11}.

\citet[708]{Fiske92} gives several examples of MP that can be compatible with the fatalistic ideology, noting that ``many writers have confused MP with asocial interactions. Some of the most egregious evils of MP are prostitution, capture and sale of people into slavery, the killing of indigenous inhabitants to open land up for economic exploitation, child labor, and colonial systems of forced labor. Mercantile wars fought for markets and sources of raw materials are also high on the list, as is the violence that is intrinsic to the businesses of drug dealing, loan sharking, and extortion." 

Moreover, \citet[38]{ClumsyVerweij11} suggests that fatalistic actors seek ``relative (rather than absolute) gain," which can be seen as an implementation of zero-sum MP (what someone gains is what someone else loses). Also, MP is present by default whenever goods are sold and bought, which can hardly be avoided in modern societies. Yet, in the village studied by \citet[44-5]{Banfield58}, consumption is kept down to a minimum. In general, agreeing on fair prices can be next to impossible in fatalism, leaving room, again, only to the asocial interaction (e.g. by setting abusive prices). In Banfield's account, landlords and tenants give up on an MP relationship out of mutual distrust (ibid., 93). Furthermore, the employer-employee relationship could be MP but instead turns into the asocial interaction (``the worker is usually cheated", ibid., 93). Let us note that the latter case could be seen as a violation of MP, which still implies an MP-type of reasoning. Finally, fatalistic MP may take the form of corruption (ibid., 102-3).

Fatalistic mutual distrust is unfavorable for CS relationships, which in principle require frequent interactions and considerable trust and commitment. In order to overcome this mutual distrust, CS relationships (especially their initiation) may take extreme forms, such as violent blood rituals. This might serve to raise the stakes in an attempt to ensure loyalty in a non-conducive environment. In general, characteristic features of fatalistic CS relationships are that they may be short-lived, unreliable and dependent on changing circumstances. Or, they may actually endure, but under the constant threat of being broken off. The ancient proverb ``the enemy of my enemy is my friend" may be an illustration of such a self-interested, contextual, temporary and dubious form of solidarity. 

It is quickly done to defect in an EM relationship, especially if it entails the reciprocal exchange of favors over time. The mostly fatalistic peasants interviewed by \citet[121]{Banfield58} consider that EM, if used, has to be applied strictly: ``Even trivial favors create an obligation and must be repaid." Such relationships are thus avoided: ``You would needlessly create an obligation which you would have to repay." An expression of EM that can be common in fatalism is in-kind retribution (tit-for-tat), a canonical example of EM that does not require any agreement from the involved parties. \citet[705]{Fiske92} mentions that ``Retaliatory feuding and vengeance are often based on EM." 

\subsection{Summary} \label{S-summary}

Our approach consists in arguing that each RM may be implemented in a way compatible with each cultural bias, and offering examples of such implementations. Importantly, this implies that an association may be found between each RM and each cultural bias. In other words, there is no one-to-one mapping between the RMs and the group-grid dimensions or the cultural biases. 

Our findings from sections \ref{S-biases-H} to \ref{S-biases-F} are summarized in table \ref{T-biases-RMs}. This table shows the four cultural biases and lists how the RMs may be implemented within each bias, based on the examples we offer in sections \ref{S-biases-H} to \ref{S-biases-F}. Let us stress that table \ref{T-biases-RMs} does not provide an exhaustive list of the various possible implementations of the RMs in each bias. This is impossible in principle, as each RM can be expressed in an indefinite number of ways. Table \ref{T-biases-RMs} illustrates and summarizes our thesis that all RMs have implementations compatible with each cultural bias.

\begin{table}[!ht]  
\begin{center}
\begin{tabular}{|c|c|c|} 
\hline
& {\bf Fatalism} & {\bf Hierarchy} \\
\hline
&{\bf CS:} unreliable ; & {\bf CS:} abiding by collective rules   \\ 
& extreme initiations & (e.g. food rituals in caste system; \\
&  (e.g. blood rituals) & intolerance of same-sex relationships) \\
& {\bf AR:} unreliable mentor-prot\'eg\'e & {\bf AR:} legitimate authority \\
& relationship; & (based on prescribed roles) \\
High grid & {\bf EM:} strict reciprocity; tit-for-tat & {\bf EM:} cooperation \\
& {\bf MP:} evil, amoral & {\bf MP:} top-down allocation \\
& {\bf Asocial:} default state & {\bf Asocial:} violation of rules; punished \\
& {\bf Null:} prevents changes & {\bf Null:} exclusion \\ 
\hline
&{\bf Individualism} & {\bf Egalitarianism} \\
\hline
&{\bf CS:} whatever involved parties & {\bf CS:} consensus decision-making;  \\ &  want for themselves; & abiding by collective rules \\
& merging companies & emphasizing absence of discrimination \\ 
& {\bf AR:} expertise-based;  & {\bf AR:} charisma-based; \\
& prestige ranking & rotating chair(wo)manship \\
Low grid & {\bf EM:} equal opportunity; & {\bf EM:} equal opportunity \emph{and} outcome; \\
& one-person-one-vote; alliance & hinders discrimination  \\ 
& {\bf MP:} contract, bargain & {\bf MP:} such that everyone benefits \\
& (individual optimization) & (e.g. barter, microcredit) \\
& {\bf Asocial:} expected;  & {\bf Asocial:} pathological;  \\
& triggers self-defense & helped toward reintegration \\
& {\bf Null:} voluntary withdrawal & {\bf Null:} exclusion \\
\hline
& Low group & High group \\
\hline
\end{tabular}
\caption[Implementations of relational models compatible with each cultural bias]{This table summarizes the examples of implementations of RMs compatible with each cultural bias given in sections \ref{S-biases-H}-\ref{S-biases-F}. This suggests that an association may be found between each RM and each bias. This table also describes how asocial and null interaction may be perceived within each cultural bias. We list RMs in the conventional order used in RMT.} \label{T-biases-RMs} 
\end{center}
\end{table}

Our approach can be positioned in PRT by drawing a parallel with the way cultural biases are characterized in case studies. Namely, a table is often given which lays out features of each bias in a relevant social context. For example, \citet[63, 88]{Olli12} lists characteristics of household organization, \citet[56-7]{ClumsyVerweij11} focuses on global governance, and \citet[35]{Coyle94} on environmental and land-use cultures. To such tables, one could add a row labeled ``relational models" and fill it with the boxes of table \ref{T-biases-RMs} (or add one row for each RM, the asocial and null interactions). 

From the RMT viewpoint, our thesis can be seen as an attempt to build on the fact that RMs are operationalized in accordance with cultural implementation rules specifying when, to whom and with regard to what they apply (\citealp[713]{Fiske92}, \citealp{Fiske2000}). Cultural biases do not specify all details about RM implementations, but they can be seen as a typology of implementation rules. Namely, each RM can be implemented in indefinitely many ways, but these ways can be categorized into the four cultural biases, according to the mutual compatibility between a given RM implementation and a cultural bias.

\section{Underlying theoretical basis}  \label{discussion}

\subsection{Different levels of analysis}

Using a number of concrete examples to support our claim, we argue above that the following connection holds between RMT and PRT: any relational model may have a number of expressions compatible with each cultural bias. A cultural bias restrains the range of congruent RM implementations, but does not preclude any RM altogether. 

We now justify this idea at a theoretical level by proposing that a relational model is attached to a relationship between two actors (that can be persons or groups), whereas a cultural bias emerges from a larger pattern of social relationships. By larger pattern, we mean a set of nodes (actors) and links (social ties), as in graph or network theory. While this representation is in fact not new, it is nonetheless problematic, as we now discuss.

The idea that cultural biases emerge from patterns of social relationship is at the basis of PRT, but it is not fully solved. \citet[11-13]{CulturalTheory90} describe what such patterns should look like for each category of the typology, but only textually. \citet[47]{Perri06} and \citet[19]{Olli12} propose one type of network for each quadrant of the grid-group diagram, but without rending explicit how the grid and group axes relate to quantities that can be measured on networks. \citet{GrossRayner85}, for their part, define measurements of both the grid and group dimensions on networks, which they illustrate graphically. However, they do not offer a generic type of network for each quadrant of the grid-group diagram. 

Missing in PRT is thus a coherent network representation, which would for example offer well-defined grid-group axes giving rise at their extreme ends to ideal structures (i.e. patterns of social relationships), relating in clear ways to cultural biases. The absence of such a representation is linked to the issue that it is actually not clear how cultural biases derive from the grid and group dimensions. The latter problem is notably reviewed by \citet[283-93]{Olli12}.

Attempting to resolve these issues in PRT is beyond the scope of the present work. For the time being, we assume that it is possible to define measures of grid and group on social networks, giving rise to certain network structures that clearly support and are supported by cultural biases. This assumption justifies the idea that cultural biases are attached to different types of networks. So far, we have not directly focused on this theoretical idea. Instead, we have defended a rather empirical consequence of this idea, namely that specific RM expressions are compatible with the ideology associated with the larger social structure that these relationships inhabit and contribute to create. 

Even if the aforementioned issues were solved in PRT, the proposition that RMs apply to single ties and cultural biases to networks would still not be straightforward, in the context of either PRT or RMT. At first sight, it may appear to overlook several facts:
\begin{enumerate}
\item PRT may be argued to apply to relationships between two actors as well; for example a ``hierarchist relationship" may be seen as a relationship whose features and actors' views are consistent with the hierarchical bias and associated high-grid, high-group pattern; 
\item RMT has been extended to describe the combinatorics of relational models through the introduction of ``metarelational models" \citep{Fiske2012}, thus going toward describing networks;
\item According to another extension of RMT, relationship regulation \citep{RaiFiske11, FiskeRaiVV}, each RM generates a distinct ``moral motive," i.e. a set of moral values, which evokes cultural biases.
\end{enumerate}

Thus, both theories have something to say about each of several levels: individual and institutional sets of values, perceptions, etc.; relationships between two individuals or groups; and structures of social networks. 

Of the above points, the simplest to address is the one about PRT possibly applying to relationships between two actors. It is indeed the case that a network (a set of nodes connected by links) can be so small as to include just a pair of nodes connected by a single link. In that sense, PRT can talk about, say, an egalitarian relationship, meaning that the constitution of this relationship supports and is supported by the principles of the egalitarian bias. According to our framework, RMT provides a more precise characterization of this relationship: namely, it can be structured in the ways given by the four relational models, implemented in ways compatible with the egalitarian bias.

The first point thus raises no contradiction to our theoretical hypothesis. The two other points are treated in the next sections, and argued not to constitute contradictions to our hypothesis either.

\subsection{Metarelational models: micro-macro vs macro-micro approach} \label{S-bottom-up}

\cite{Fiske2012} studies the combinatorics of social relationships, which examines how  combinations of relationships may be implied, required or prohibited. For instance, relationships between parent and children may inform relationships between children. This situation can be represented as a three-node network, composed of one parent and two children, with the two parent-child links entailing or prohibiting a child-child relationship. Moreover, in a dyad, two kinds of relationships may require or preclude each other. For example, AR and MP may typically be combined in the relationship between an employer and employee \citep[10]{Fiske2012}. 

\cite{Fiske2012} calls these combinations of relationships ``metarelational models" (MeRMs) and defines six elementary MeRMs occurring either in dyads or triads. MeRMs are thus represented as two- or three-node networks. Fiske also offers two combination mechanisms, recursive embedding and temporal dynamics, which correspond to integrating (or summing) over people and time, respectively. Recursive embedding allows the construction of larger social networks, using the elementary MeRMs as building blocks.
 
The framework we introduce in this chapter connects PRT and RMT in a macro-micro manner. That is, we start from the cultural biases, and derive in a deductive manner the forms that the RMs can take within the biases. Understanding inductively the micro-macro process of how relationships aggregate to create a larger social structure associated with a cultural bias is a different problem. In order to solve the question of the connection of PRT and RMT in this direction, one may have to find associations between the MeRMs and PRT's patterns of social relationships associated with cultural biases. As discussed above, this would require initially adopting a network-like representation for each quadrant in the grid-group diagram. 

It may be relevant to note that the latter process is an instance of the micro-macro upscaling problem, which is fundamental in physics, material sciences, biology, economics, sociology, traffic flow, and so on \citep{Anderson72,Wilson79,Sornette2004}. The micro-macro problem is not solved in general. In fact, it is solved in general terms only in the statistical physics of critical phase transitions at equilibrium, through the theory of the renormalization group \citep{Wilson79}. 

In the present case, the analysis is made even more difficult by the fact that the whole system is regulated by micro-macro and macro-micro feedback loops, i.e. cycles going both ways between the micro (individual) and the macro (social structure and culture) levels. This is another way to say that culture influences individuals and relationships at the same time as individuals and relationships are the constituents of culture.

In summary, the elementary MeRMs may be seen as a first step toward solving the micro-macro upscaling problem in the context of social relationships. Our framework, on the other hand, emphasizes the macro-micro process connecting PRT and RMT.

\subsection{Morality: relationship regulation and ``virtuous violence"}

In recent years, RMT was extended into a theory of relationship regulation \citep{RaiFiske11} and ``virtuous violence" \citep{FiskeRaiVV}. This recent extension fully develops ideas expressed earlier by \citet[130-3]{FiskeBook}.

According to relationship regulation theory, each RM generates a distinct \emph{moral motive}, respectively unity (CS), hierarchy (AR), equality (EM) and proportionality (MP). The related principle of virtuous violence holds that much or most violence derives from these fundamental motives and may thus be perceived as morally correct. In a nutshell, violence may be seen as rightly inflicted when it is perceived as protecting the group (unity), responding to authorized orders (hierarchy), maintaining balance (equality) or respecting proportionality in relevant social relationships (proportionality).

The moral motives, being sets of values (which do not apply only to violence), may seem to identify with the cultural biases of PRT. In fact, we argue that moral motives relate to cultural biases in the same way as relational models. That is, each moral motive can be activated within each cultural bias, being tied to a relational model whose specific expression is compatible with the cultural bias.

For example, the moral motive of equality can be pursued in a variety of different contexts. Ensuring equality of opportunity is likely to be seen as an obligation in both egalitarianism and individualism; however, guaranteeing equality of outcome may be seen as compulsory in egalitarianism, and at the same time undesirable or even morally wrong in individualism. The same moral motive thus applies to different social aspects depending on the cultural biases within which it is enacted.

The latter reasoning is no different than the ones carried out in sections \ref{S-biases-H}-\ref{S-biases-F}. Overall, the moral motives, being tied to social relationships, follow the same rules as relational models, relatively to cultural biases. Namely, thinking simultaneously in terms of moral motives and cultural biases allows one to understand more precisely how moral values can be enacted within cultural contexts.

Interestingly, \citet[67]{RaiFiske11} note how cultural settings valuing freedom are associated with restriction of AR and expansion of MP: 

``In some cultures, freedom - autonomy, independence - is a core moral and political value. [...] Future research should explore how restriction of AR has combined by expansion of MP to form the integrated psychocultural construct of freedom. In particular, how does `freedom' interact with other moral motives to restrict the scope of some social-relational (and consequently moral) obligations, such that beyond these boundaries people can and \emph{should} pursue their own interests without regard to the needs and desires of others and any attempts to forcibly impose social-relational obligations are regarded as illegitimate."

It appears that this quote makes a point similar to the development we make in section \ref{S-biases-I}, where we discuss how PRT's individualism (pursuing freedom) selects compatible implementations of RMs (of which AR and MP); by extension, this also concerns moral motives.

\citet[67]{RaiFiske11} also note the coupling of the hierarchy and unity moral motives in religious contexts, in line with the ideas expressed in section \ref{S-biases-H} concerning the coexistence of AR and CS in PRT's hierarchical setting: ``[...] Unity violations may sometimes become coupled with motives for Hierarchy in religiously based concerns where individuals cast Unity-violating acts as disobedience to God's will or injury to God's flock." 

Morality is also a concern on the PRT side. For example, \cite{ChaiWildavsky94} and \cite{LockhartFranzwa94} examine the relationship between morality, violence and culture. Moreover, \cite{Bruce13} introduces a model unifying PRT, Jonathan Haidt's moral foundations theory \citep{Haidt07} and Richard Shweder's tripartite theory of morality \citep{Schwederetal97}. This suggests that our framework could be used in future research to elaborate on how social-relational contexts interact with morality, and on possible connections between different theories of morality.

\section{Review of previous studies and suggestions for empirical research} \label{S-studies}

\subsection{Vodosek (2009): RMT and Triandis' constructs (1995)}

Our framework sheds light on the attempt carried out by \cite{Vodosek09} to test hypotheses expressed by \citet[50-1]{Triandis95} and \cite{Triandis98} concerning the relationship between relational models and the constructs introduced by \citet{Triandis95,Triandis96}. Triandis defines four types of social settings, namely vertical individualism, horizontal individualism, vertical collectivism and horizontal collectivism. The correspondence between cultural biases and these constructs is suggested by \citet[87-9]{Verweij14Disagree}. In line with this, we identify vertical collectivism with hierarchy, horizontal collectivism with egalitarianism, horizontal individualism with individualism, and vertical individualism with fatalism. 

\citet[50-1]{Triandis95} and \cite{Triandis98} hypothesize that vertical relations correspond to AR, horizontal ones to EM, individualism to MP and collectivism to CS. This results in the association of two RMs with each construct, as shown in table \ref{T-Vodosek}. This table also shows the results of Vodosek's test of these hypotheses, under the form of presence or absence of association.

\begin{table}[!ht]  
\begin{center}
\begin{tabular}{|c|c|}
\hline
{\bf Vertical individualism} & {\bf Vertical collectivism} \\
{\bf (Fatalism)} & {\bf (Hierarchy)} \\
\hline
AR: yes & AR: yes  \\ 
MP: no & CS: yes \\
\hline
{\bf Horizontal individualism} & {\bf Horizontal collectivism} \\
{\bf (Individualism)} & {\bf (Egalitarianism)} \\
\hline
EM: no & EM: yes \\
MP: no & CS: yes \\
\hline
\end{tabular}
\caption[Triandis' hypotheses (1995) and Vodosek's results (2009)]{This table summarizes the hypotheses expressed by \citet[50-1]{Triandis95} and \cite{Triandis98}, and shows the results of Vodosek's tests of these hypotheses (2009). We write ``yes" when a significant association is found by Vodosek (confirming a hypothesis expressed by Triandis et al.) and ``no" otherwise. We assume that vertical individualism corresponds to fatalism, vertical collectivism to hierarchy, horizontal individualism to individualism, and horizontal collectivism to egalitarianism.} \label{T-Vodosek} 
\end{center}
\end{table}

Our framework predicts that it is possible to find significant associations between any RM and each cultural bias, as long as the expression of the RM is compatible with the bias. For example, CS may be associated with hierarchy, but not if it is expressed as consensus decision making, an implementation of CS that happens to be incompatible with the hierarchical bias.

It is thus relevant to explain the few absences of association shown in table \ref{T-Vodosek}, which all concern EM or MP. The items used by \citet[128]{Vodosek09} for these two RMs are the following.
\begin{enumerate}
\item[EM1:] Group members typically divide things up into shares that are the same size.
\item[EM2:] Group members often take turns doing things.
\item[EM3:] When group members work together, they usually split the work evenly.
\item[EM4:] Group members make sure that the group's workload is shared equally.
\item[EM5:] The group makes decisions by a simple majority vote.
\item[MP1:] Group members calculate what their payoffs are in this group and act accordingly.
\item[MP2:] Group members divide things up according to how much they have paid or contributed.
\item[MP3:] Group members make decisions according to the ratio of the benefits they get and the costs to them.
\item[MP4:] Group members choose to participate in the group when it is worth their while to do so.
\end{enumerate}

Overall, the absence of association of EM and MP with low group biases may be due to all EM and MP items mentioning ``group" or ``group members," while the notion of by-default group membership is contrary to low group settings. 

More precisely, regarding EM items, the main issue may be that, in individualism, people are not bound to apply EM to all task or resource distribution problems. They do so only if it makes sense in a given situation and in the view of the individuals involved. Individualistic task distribution is thus rather skill-based. Based on interviews in an individualistic household, \citet[219]{Olli12} notes that ``The division of housework is quite flexible [...] Equality is not created by spending an equal number of hours on household tasks, [...] tasks are allocated to the person who has the better skills and more time to do them." 

Item EM5, expressing the one person, one vote principle, may evoke individualism, but the formulation ``The group makes," evoking collectivism, may have induced participants to reject this item in association with individualism. Finally, some of these items imply a notion of collective supervision (e.g. EM4) contrary to low group. 

In relation to individualism, a common issue with MP items is that they all evoke collective action or decision. Even MP1 and MP3, which at first sight should appeal to individualistic self-interest, still imply that a payoff results from acting specifically in or with the group. Also, the calculation aspect present in MP1-4 may imply that such calculations constitute the only appropriate strategy. In fact, individualistic actors feel free to act based on their personal preferences, without necessarily justifying them on the basis of a cost/benefit ratio.

Overall, it is thus possible that the framing of the EM and MP items induced participants to reject any association between EM, MP, and individualism. Even with due consideration of this framing effect, one may still have difficulty finding any striking association between any RM and the individualistic bias because of the relative RM-flexibility of individualists, as argued in section \ref{S-biases-I}. Individualistic actors constitute their relationships along the lines of the RMs, but feel free to pick the RM and the specific implementation thereof that suit them best individually (as long as these also suit the other directly involved parties). Designing generic RM items that happen to suit individualists in general may thus be challenging. We come back to this point in section \ref{S-research}.

As in individualism, the absence of association between MP and fatalism may be due to the mention of ``group members" and the implication that these members agree on shared MP schemes. This is in principle incompatible with fatalists' isolation and lack of coordination. We argue in section \ref{S-biases-F} that fatalistic MP is mostly amoral, or appears under the form of failed (violated) MP.

Vodosek's results aggregate items relating to the same RM, but it is interesting to note that even where associations were found, some specific items should not correlate with the cultural setting. For example, CS is generally found to correlate with hierarchy, but one of the CS items is ``The group makes decisions together by consensus" \citep[128]{Vodosek09}. This is not a hierarchical view; it rather evokes the egalitarian bias. We thus hypothesize that this particular CS item was not associated with vertical collectivism. 

Our analysis thus helps explain Vodosek's results, in particular the absence of association between EM, MP and low-group cultural biases. Based on our framework, we suggest that associations can potentially be found between each RM and each cultural bias, but that in practice this depends on the specific formulation of the RMs.

\subsection{Verweij (2007): critique of RMT} 

\citet{Verweij07} reviews the similarities of PRT and RMT, as well as their respective shortcomings and strengths. Particularly interesting in the light of our framework is his suggestion that the relational models appear to be ``somewhat overlapping" (p. 10), which he illustrates with the following examples.

``[...] in competitive markets (i.e., transactions that resemble the `ideal markets' of neo-classical economics), actors are certainly `equal, but distinct peers': none of them has more market power than any of the others, and none of them can hope to earn a higher wage, or make a larger profit, than the others. Yet, these markets not only fulfill the conditions of equality matching, but also of market pricing." (p. 10)

In our approach, this situation appears to correspond to an instance of EM (equal opportunity) within an individualistic setting (competitive markets). It is no contradiction that such markets simultaneously function on the basis of MP in (many) other respects; on the contrary, this idea is supported by our framework. This case is particularly reminiscent of the examples given in section \ref{S-biases-I} of entire processes guided by a given cultural bias and encompassing several distinct RMs at different stages. The other points raised by Verweij happen to follow a similar logic.

``Similarly, in many Japanese markets [...], each major company is closely related - in an almost paternalistic way - to a set of smaller competitors, suppliers and retail firms. When profits and prices fall or cheaper alternatives become available, the large company at the centre of the network will not abandon the peripheral enterprises, but instead offer them advice as to how to reorganise their companies and even forms of temporary financial support. These practices constitute authority ranking and market pricing at the same time." (p. 10)

The organization of Japanese markets in this example appears to be highly hierarchical and built on the basis of AR relationships, while the intervention of the larger company follows an MP logic. This is a typical example of an MP-based resource allocation determined in a top-down hierarchical manner (section \ref{S-biases-H}). Need-based advice and support may even correspond to CS under a form consistent with hierarchical implementation rules.

Finally, mentioning the fatalistic society described by \citet{Banfield58}, Verweij notes the folllowing:

``[...] a particular form of equality matching is also prevalent when social relations are characterised by extreme levels of inter-personal distrust and jealousy. [...] no cooperation will take place if anyone benefits more from the collective effort than the others. Fiske calls such relations `asocial relationships' [...] Still, a perverse form of equality matching is characteristic of such null relationships." (p. 10) 

This is actually compatible with the implementation of strict or ``perverse" EM in a fatalistic setting otherwise favoring asocial and null interactions, as we proposed in section \ref{S-biases-F}.

Our approach is thus able to answer Verweij's points by reframing his examples in terms of distinct RMs coexisting with each other, but not overlapping, within larger social settings characterized by PRT's cultural biases.

\subsection{Brito et al. (2011): RMT and families, friendships and work} \label{S-Brito}

Our approach may also provide an interpretation to the findings of another empirical study. \citet{Brito11} examine the question of which RMs are present within different social contexts. They find that relationships within families are mainly of the CS and AR types, friendships show all RMs except for AR, and work organizations correlate with MP and AR. 

In the light of the study of household organization performed by \cite{Olli12}, and of PRT case studies in general, we would expect families, work organizations and friendships to differ from each other in ways that may be described by cultural biases. Yet, Brito et al. do not evaluate the cultural bias of social contexts from the viewpoint of PRT or any related theory. They appear to assume that in terms of favored RMs, and in the context of their study, all families are similar, as well as all work settings, and so on. In the context of our framework, this limits the interpretation that we can make of their findings, but their study presents interesting points, among which are the following.

One of Brito et al.'s findings is that EM is generally correlated with CS (p. 423), which would make sense in our framework if the cultural bias happens to be egalitarian or hierarchical. Indeed, in these cases, the same actors who interact on the basis of one of these two RMs are more likely to interact on the basis of the other as well.

Brito et al. encountered difficulties with a few MP items that correlated with each other, but badly with other MP items (p. 417). One of these problematic items is: ``What you get from this person is directly proportional to how much you give them" (p. 431). Interestingly, this statement can fairly easily be imagined to be true of love, violence, attention, time, respect, effort, (im)politeness, and so on, and be valid in many contexts, including home, work and friendships. What it expresses is generally a mechanism of positive feedback, which can apply to many different behaviors or emotions. Hence, it seems rather devoid of content and potentially correlated with any social context. From the perspective offered by our framework, we would thus argue that it is actually a very appropriate item for MP: it can be expected to correlate with any social context, in line with the view we express in the present chapter that all RMs may be present in any cultural bias. However, Brito et al's reached an opposite conclusion: in fact, they removed this item from their analysis (p. 417). In contrast, the MP items that they kept mention payment, payoff, costs, rationality, or regulations, which all evoke the context of work more than anything else. It is thus no surprise that these other items correlated well with work, and only partly with friendships.

\subsection{Realo et al. (2004): RMT and collectivism}

\cite{Realo04} examine associations between RMs and collectivist attitudes, understood in the sense of \cite{Triandis95,Triandis96} and \citet{Triandis98}. They look for associations between RMs and different contexts, namely family (familism), peers (companionship) and society (patriotism). Realo et al.'s approach and conclusion are very similar to our own. On a theoretical basis, Realo et al. express the following opinions, questioning the possibility of a one-to-one mapping between RMs and cultural settings in the same way we do:

``[...] a simple one-to-one link between Fiske's elementary forms of sociality and collectivism-individualism can be questioned [...] one-to-one relationships between cultural syndromes and elementary constituents of human relations are unrealistic: relational models theory is not a taxonomy of cultures. Instead, the models are abstract and open [...]" (p. 784)
 
``[...] we expect that individuals scoring high on collectivism will use the whole range of elementary forms of sociality to organize their different relationships, or even the different aspects of the same relationship. We believe that there is not necessarily a one-to-one association between the major cultural dimensions and the elementary models of social relations." (p. 785)
 
Realo et al.'s thesis is supported by their results. They do not give the twenty-three items they used to measure the RMs, so we cannot attempt any precise interpretation. Overall, they state that:

``[...] the relationships between the use of the relational models and the collectivist attitudes were far from strict one-to-one associations. It is quite clear that people scoring low on collectivism do not always use MP and not in all situations; neither are `collectivists' rigidly fastened to the use of CS, not even in their family relations. In most of the hypothetical situations, both people scoring low and high on collectivism tended to use all four models as dictated by general cultural norms, the logic of the situation, and their personal history." (p. 789)

This is clearly in line with the view we develop in this chapter.

These comments on previous works suggest further possible ways to test our hypotheses and design empirical research, which is the topic of the next section.

\subsection{Further testing of our hypotheses: empirical research design} \label{S-research}

Our framework suggests that associations may potentially be found between each cultural bias and each RM. In sections \ref{S-biases-H}-\ref{S-biases-F}, we support this thesis by providing a number of examples mostly drawn from PRT case studies, but also from the RMT literature. The fact that we found various supporting illustrations (and no contradicting ones) can be seen as a first successful test of our hypotheses. Additional tests could be carried out as follows. 

First, one may perform case studies (as in PRT) and identify manifestations of cultural biases and RMs at the same time. In contrast, in the present work, we typically identify RMs a posteriori in PRT case studies completed by others who presumably did not have our framework in mind. When attempting to identify RMs in live situations, the constitutive media for each RM given by \citet[64]{FiskeConstitMed04} may be useful. For example, sharing comestibles, dancing and body contact are some of the most emotionally evocative constitutive media for CS; icons or people set above or in front of others are constitutive media for AR, and so on. 

Incidentally, this approach may bring insights into PRT case studies. By clarifying counterintuitive associations, it may help determine with greater confidence the cultural bias(es) at work within a social context. For example, it may help make sense of individualistic or hierarchical actors insisting on equality of, respectively, opportunity, or same-rank people, for example. Such expressions of EM are coherent with the individualism or hierarchical bias, respectively; they are not counter indicators thereof, and do not necessarily amount to actors borrowing the rhetoric of egalitarianism. This principle also holds for other RMs in other biases, such as implementations of MP in egalitarian settings, and so on.

Second, one may perform survey-based analyses similar to the ones carried out by \cite{Vodosek09, Brito11} and \cite{Realo04}. That is, survey questions could be aimed at assessing both cultural biases and relational models in given social contexts and relationships, in order to test whether associations correspond to those of table \ref{T-biases-RMs}. 

This approach, however, is not guaranteed to succeed. For one thing, reservations have been expressed regarding survey-based tests of PRT (\citealp{Verweij11HowToTest}, \citealp[208]{Olli12}, \citealp{Verweij14Disagree}). For another, it may be challenging to find sufficiently generic items evoking RMs compatible with cultural biases. For example, observing elaborate food rituals whose details heavily depend on the participants' status in a hierarchical system can be interpreted as an expression of ``hierarchical CS," i.e. an implementation of CS compatible with the hierarchical cultural bias. Yet, finding a generic survey item evoking hierarchical CS may be problematic. There are indefinitely many possible expressions of each RM, and different hierarchical social systems may use different specific implementations of CS. In general, one can expect that hierarchical CS relationships are respectful of and determined by the hierarchy; differently said, in a hierarchical setting, group pressure and prescribed roles constrain CS relationships. But expressing this abstract principle in a survey may not bring the intended results, because people are not necessarily aware in these terms of their own social relationships.

Therefore, case study research might be more appropriate than survey-based methods in order to further test our hypotheses. Nevertheless, if one wants to go about surveys, one should test for $24$ constructs: ($4$ RMs + 1 asocial + 1 null) $\times$ $4$ cultural biases, and define 5 or 6 items for each construct. Starting points to define questions could be \citet{Dake91}, \citet{BoyleCoughlin94}, \citet{Vodosek09} or \citet{Brito11} and \citet[308]{Olli12}. For example, the item ``I support a tax shift so that burden falls more heavily on corporations and people with large incomes" \citep[69]{Dake91} may test for ``egalitarian MP," and ``Everyone should have an equal chance to succeed and fail without government interference" \citep[308]{Olli12} may test for ``individualistic EM."

\section{Conclusion}

Our framework proposes that each relational model may be expressed in a way compatible with each cultural bias. A cultural bias restrains the range of congruent RM implementations, but does not preclude any RM altogether.  We support this idea by offering examples drawn from the literature of both RMT and PRT. This implies that associations may be found between any RM and each cultural bias, and that there is no one-to-one mapping between RMT and PRT. 

At a theoretical level, our thesis is based on the hypothesis that PRT and RMT apply to different levels of analysis: a relational model is attached to a social relationship (i.e. a tie between two actors), whereas a cultural bias emerges from a larger pattern of social relationships (i.e. a social network). From this representation derives the idea that social relationships instantiating relational models are compatible with the ideology associated with the larger social structure they inhabit, support and constitute. Future work shall address this theoretical representation, in particular by offering network types for each cultural bias. 

Our framework can inform empirical research, by clarifying the social relationships that can be expected in different social contexts assessed through the typology of PRT. It also helps interpret previous (one-to-one) reconciliation attempts, and opens up paths for future research. Overall, by attempting to connect two theories of constrained relativism, our hope is to strengthen this movement and facilitate future research in this domain.

\section*{Acknowledgments}

Marco Verweij and Alan P. Fiske gave much appreciated feedbacks on an earlier version of this work. These ideas benefited from discussions with Alexandru Babeanu and Catherine Alexander about the respective domains of applicability of RMT and PRT. 

\bibliography{RMT-CT}  
\bibliographystyle{ametsoc} 

\end{document}